\documentclass[10pt,letterpaper]{article}
\usepackage{opex3}
\usepackage{upgreek}

\begin{document}

\title{Three-dimensional direct laser writing inspired by \mbox{stimulated-emission-depletion microscopy}}

\author{Joachim Fischer$^{*}$ and Martin Wegener}

\address{DFG-Center for Functional Nanostructures (CFN), Institut f\"ur Angewandte Physik, and Institut f\"ur Nanotechnologie, Karlsruhe Institute of Technology (KIT), \\D-76128 Karlsruhe, Germany}

\email{$^{*}$joachim.fischer@kit.edu} 



\begin{abstract} Three-dimensional direct laser writing has become a well established, versatile, widespread, and even readily commercially available ``workhorse'' of nano- and micro-technology. However, its lateral and axial spatial resolution is inherently governed by Abbe's diffraction limitation -- analogous to optical microscopy. In microscopy, stimulated-emission-depletion approaches have lately circumvented Abbe's barrier and lateral resolutions down to 5.6\,nm using visible light have been achieved. In this paper, after very briefly reviewing our previous efforts with respect to translating this success in optical microscopy to optical lithography, we present our latest results regarding resolution improvement in the lateral as well as in the much more relevant axial direction. The structures presented in this paper set a new resolution-benchmark for next-generation direct-laser-writing optical lithography. In particular, we break the lateral and the axial Abbe criterion for the first time.\end{abstract}

\ocis{(350.3390) Laser materials processing; (350.3450) Laser-induced chemistry; (220.4241) Nanostructure fabrication; (160.1245) Artificially engineered materials.} 


\section{Introduction}
Direct laser writing (DLW) \cite{Misawa,Kawata,Minggu,Deubel} has become a workhorse for fabricating complex three-dimensional (3D) micro-structures or even nano-structures. In essence, a laser is very tightly focused into the volume of a photoresist (usually on a glass substrate). By using two-photon absorption or other nonlinearities, the photoresist is only sufficiently exposed within the focal volume. This tiny volume element is often referred to as the ``voxel'' in analogy to the pixel (picture element). By scanning this voxel with respect to the photoresist, essentially arbitrary three-dimensional objects can be fabricated. We have recently reviewed corresponding examples from photonics \cite{Georg} (also see references therein). This review \cite{Georg} also compares DLW with other 3D fabrication approaches. Particularly complex photonic structures recently made by DLW in our group comprise 3D icosahedral quasicrystals \cite{Alexandra}, 3D rhombicuboctahedral quasicrystals \cite{Alexandra2}, waveguides in 3D photonic-band-gap structures \cite{Isabelle}, gold helix-metamaterials \cite{Justyna}, and 3D polarization-independent invisibility cloaks operating at telecom wavelengths \cite{Tolga}. We have recently also realized biologically meaningful one-component \cite{Klein} and two-component \cite{Beni} 3D scaffolds for cell growth studies - possible future ``designer Petri dishes''.

Compared to electron-beam lithography or deep UV lithography, the resolution of DLW is an issue. Potential applications like optical data storage with enhanced data density or three-dimensional photonic metamaterials clearly demand for higher resolution.
In general, the resolution of an optical system is limited by the laws of diffraction. In optical microscopy, Ernst Abbe found the lateral resolution to be $d_{\rm lateral}=\lambda/(2\rm NA)$ with the numerical aperture NA. Two simultaneously emitting point sources separated by a smaller distance can not be distinguished.

Although the resolution of DLW optical lithography is often associated with the terms ``sub-wavelength'' or even ``sub-diffraction-limited'', its lateral and axial resolution are still fundamentally limited by Abbe's law. The critical distance between two point sources can be translated to the minimum distance between two point exposures. Clearly, the use of two-photon absorption shifts this resolution limit by a factor of $\sqrt{2}$ (assuming Gaussian profiles), as the exposure dose is proportional to the squared intensity rather than the intensity itself.

The resulting width of lithographic lines or point exposures can be further reduced. As this quantity has no equivalent in optical microscopy, Abbe does not make a prediction here. By putting the maximum of the spatial laser intensity close to the polymerization threshold, only the innermost region of the exposure volume is sufficiently exposed. In this way, lateral feature dimensions around 80\,nm (\textit{i.e.}, about ten times smaller than the exposure wavelength of typically $0.8\,\rm \upmu m$)  are achievable with a commercial system \cite{nanoscribe} in suitable photoresists. However, going closer to the threshold makes the fabrication process more sensitive to laser fluctuations. Moreover, the monomer-conversion is negatively affected, leading to a low cross-linking density and hence a lack of mechanical stability.

Being able to create features with sub-wavelength dimensions one might intuitively think that also sub-diffraction-limited distances are possible. As DLW is a serial process this could indeed be possible, as the Abbe criterion only truly holds for simultaneous exposure (corresponding to simultaneously emitting sources in microscopy). If the resist regions directly adjacent to an already polymerized line (where the exposure intensity was below threshold) can regenerate due to diffusion exchange another line could be polymerized directly next to the first one. In this case (where the resist has no memory for sub-threshold exposures), the center-to-center distance would only be limited by the line width. However, this turns out to be very difficult and, in fact, to the best of our knowledge, not a single  publication has shown two adjacent yet separated features with a distance below the simple Abbe condition (of course, modified for the two-photon absorption). This condition states the smallest possible center-to-center distance is $d_{\rm lateral}=\lambda/(2{\rm NA}\cdot \sqrt{2})\approx 205\,\rm nm$  for the lateral direction, where $\rm NA=1.4$ is the numerical aperture and $\lambda =810\,\rm nm$ is the free-space exposure wavelength used in our setup. Two corresponding squared focal intensity profiles, spatially shifted by this distance and added up, result in a flat-top distribution, so that the two peaks cannot be separated by any thresholding mechanism. The axial resolution in 3D DLW optical lithography (with NA=1.4 and for a photoresist with refractive index around 1.5) is at least 2.5 times worse than the lateral resolution. In practice, resulting polymer features often show a significantly stronger elongation up to a factor of 5.5, despite using $\rm NA \ge 1.4$ \cite{Fourkas, scalinglaws}. The Abbe formula for the axial direction becomes $d_{\rm axial}=2.5\cdot d_{\rm lateral}\approx 510\,\rm nm$. This simple reasoning has also been confirmed by full vectorial calculations using Debye theory (not shown). 

We conclude that -- despite the serial exposure scheme -- the spatial resolution in DLW still seems to be ultimately limited by good-old optical diffraction and many years of research have raised little hope that periodicities or line widths on the order of 10\,nm may ever be reachable by regular DLW. Clearly, all above potential applications of DLW demand for smaller periodicities and hardly any application can make use of reduced feature sizes without reduced distances. To be honest, the worst spatial resolution actually determines the overall resolution if arbitrary complex 3D structures are aimed at. In this light, with minimum axial feature sizes of roughly 250\,nm 3D DLW has not even yet become a ``nano-technology'' but is rather still a ``micro-technology'', as the borderline between the two is commonly set to be at 100\,nm. 

Thus, novel approaches enabling systematic future improvements of the spatial resolution of 3D DLW are highly desirable. In this paper, we outline our approach that is inspired by stimulated-emission-depletion optical microscopy (Section 2). We very briefly review previously published work and emphasize our latest (unpublished) best results on test structures in Section 3 and Section 4.

\section{The Stimulated-Emission-Depletion Approach}
Throughout the last decade, Stefan Hell's idea of stimulated-emission-depletion (STED) fluorescence microscopy has revolutionized optical microscopy \cite{Hell1, Hell2, Hell3, Hell4, Hell5, Hell6} with important implications in biology \cite{Hell7}. World-record lateral resolutions down to 5.6\,nm using visible light and nitrogen-vacancy centers in diamond have been reported by Hell's group \cite{Hell8}. It is obviously interesting to translate this success of optical microscopy to optical lithography. This possibility was already mentioned by Stefan Hell in 2000 \cite{Hell2}, but only quite recently first successful experimental attempts in this direction have been published by other groups using single-photon (rather than two-photon) excitation \cite{McLeod} or a one-color scheme \cite{Fourkas}, and by our group using a two-photon two-color scheme \cite{STEDDLW,OL_Cloak}. All groups have demonstrated impressive features-size reductions \cite{McLeod,Fourkas,STEDDLW}. However, to date there has been no demonstration of sub-diffraction-limited resolution (like discussed in Section 1). Furthermore, it is still unclear whether the approaches can truly extend the capabilities of regular state-of-the-art DLW in three dimensions or whether they might be limited by some inherent drawbacks, like, \textit{e.g.}, stronger abberation sensitivity.

The key idea in all these approaches is to apply a second laser mode that locally and reversibly disables the resist polymerization. In principle, this additional switch allows to reduce the effective polymerization volume way below the usual diffraction-governed volume. However, the underlying processes of the approaches in Ref. \cite{McLeod} and \cite{Fourkas} have been assigned to be distinct from stimulated emission by the authors. Exploring stimulated emission as a possibly superior depletion mechanism has been our strategy from the start \cite{STEDDLW,OL_fs}. Let us briefly repeat this strategy and discuss competing (hence, potentially resolution-limiting) processes as well as alternative depletion channels.

\begin{figure}[htb]
\centering\includegraphics[width=13cm]{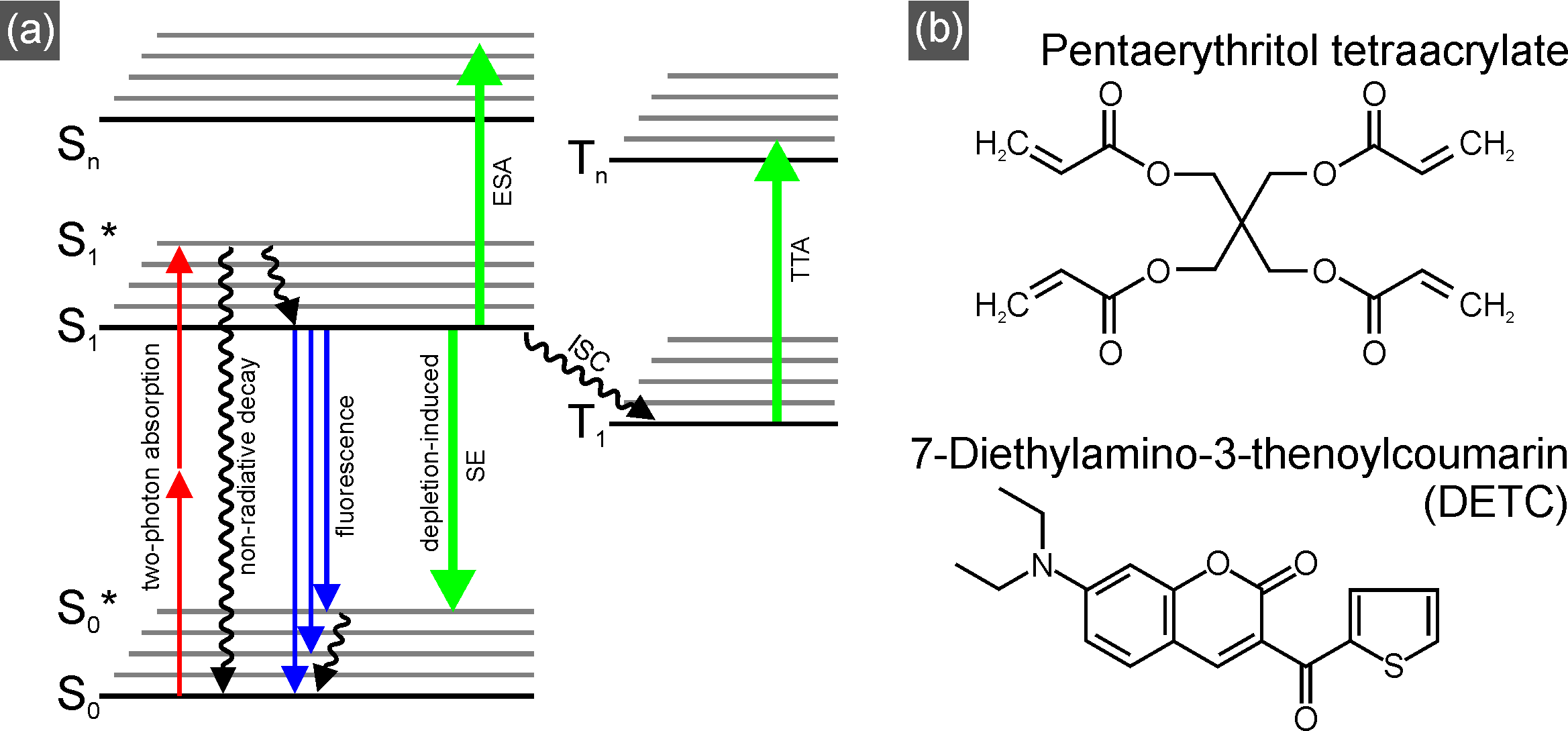}
\caption{(a) Energy level scheme with transitions in a photoinitiator molecule for stimulated-emission-depletion (STED) direct-laser-writing (DLW) optical lithography. SE: stimulated emission, ESA: excited-state absorption, TTA: triplet-triplet absorption, ISC: inter-system crossing. (b) Ingredients of our current photoresist: pentaerythritol tetraacrylate (monomer) and 7-diethylamino-3-thenoylcoumarin (photoinitiator).}
\end{figure}

The relevant states and transitions of a photoinitiator molecule are schematically shown in Fig. 1(a). The excitation (or writing) laser excites photoinitiator molecules from their ground states to an electronically and vibrationally excited level $\rm{{S_1}^*}$ \textit{via} two-photon absorption. Some molecules may directly decay back to the ground-state non-radiatively. However, the majority of molecules will relax to an intermediate state $\rm{{S_1}}$ from where they can either fluoresce or undergo inter-system crossing (ISC) to the triplet state $\rm{{T_1}}$. From the $\rm{{T_1}}$, a chemical reaction may be initiated in the photoresist. The idea of STED is to bring the molecules from the intermediate state $\rm{{S_1}}$ back to the ground state $\rm{{S_0}}$ \textit{via} stimulated emission (SE) induced by a second laser of a different color. We will call this laser the depletion laser. 

When exposing the excited molecules to the depletion laser, SE competes with excitation into yet higher-energy levels \textit{via} excited-state absorption from $\rm{{S_1}}$. Moreover, the depletion laser can be absorbed by molecules that are already in their triplet state or have already formed radicals. As highly excited molecules are often very reactive and as the unwanted absorption of the depletion beam injects additional energy into the system, such competing absorptions might enhance polymerization rates and hence counteract the depletion. This undesired cross-talk between deactivation and initiation would limit the expected potential resolution improvement and could lead to a saturating resolution above a certain optimum depletion power. 

Even if one transient absorption channel would lead to an effective depletion of the polymerization (\textit{e.g.}, due to non-radiative relaxation from a highly excited state or due to the formation of non-initiating fragments) this effect would only have limited potential for the fabrication of structures beyond the diffraction limit, as it could likely be slow, irreversible, temperature-dependent or even be mediated by a temperature increase in the resist. In contrast to potentially heat-injecting absorption-based depletion mechanisms, SE even ejects most of the excitation energy \textit{via} the generated photon. Furthermore, SE is known to be very fast and fully reversible. 

We conclude that stimulated emission is the preferable depletion mechanism in STED-inspired lithography and that transient absorptions from excited singlet states, triplet states or radicals should be avoided.

Finding a photoinitiator suitable for stimulated emission depletion is very difficult \cite{Fourkas}. To efficiently deplete a molecules excitation \textit{via} SE, a large oscillator strength is required for the $\rm{{S_1}}$-$\rm{{S_0}}$ transition. Moreover, a sufficiently large excited-state lifetime is required for the interaction. Together, one requests reasonably high fluorescence quantum efficiencies. Unfortunately, normal photoinitiator molecules are designed for fast and efficient ISC and therefore exhibit negligibly small quantum efficiencies. For example, common photoinitiator molecules like, \textit{e.g.}, Irgacure 369, Irgacure 819, Darocur TPO, Irgacure 1800, or a Michler's Ketone derivative exhibit efficiencies of $<0.1\%$, $\approx  0.2\%$, $\approx 0.3\%$, $< 0.1\%$, and $\approx 0.3\%$, respectively \cite{STEDDLW}.

Thus, our search for suitable STED-DLW photoinitiators from a plethora of possibilities has been guided by looking for larger quantum efficiencies and, of course, by looking for optical transition frequencies that are compatible with readily available laser systems. So far, we have identified two photoinitiators offering a polymerization-deactivation mechanism, namely isopropylthioxanthone (ITX) \cite{STEDDLW} and 7-diethylamino-3-thenoylcoumarin (DETC) \cite{OL_Cloak}. DETC is shown in Fig. 1(b). We have measured \cite{STEDDLW} a fluorescence quantum efficiency of $15\%$ for ITX and $3\%$ for DETC in ethanol solution. We have also measured \cite{OL_fs} molar extinction coefficients of 930\,L/mol/cm for ITX and 40550\,L/mol/cm for DETC. The latter is comparable to that of state-of-the-art green-emitting fluorescent dyes (\textit{e.g.}, Atto 425 with 45000\,L/mol/cm). For comparison, usual photoinitiator molecules like, \textit{e.g.}, Irgacure 907, have only values of around 100\,L/mol/cm for their $\rm{{S_0}}$-$\rm{{S_1}}$ transition, while their $\rm{{S_0}}$-$\rm{{S_n}}$ transitions are often undesirably strong.

By performing femtosecond pump-probe spectroscopy on ITX and DETC photoinitiator molecules in ethanol solution \cite{OL_fs}, we have unambiguously found that stimulated emission actually overwhelms excited-state absorption from the $\rm{{S_1}}$ state (see Fig. 1(a)) at 532-nm optical wavelength for the case of DETC, whereas the opposite holds true for ITX. On this basis, DETC is our present best photoinitiator for use in 3D STED-DLW. 

Using the corresponding photoresist \cite{OL_Cloak}, we have indeed recently succeeded in miniaturizing our 2010 3D invisibility cloak \cite{Tolga} -- which employed state-of-the-art regular 3D DLW -- by more than a factor of two in all three spatial dimensions such that polarization-insensitive invisibility cloaking at visible frequencies has become experimental reality \cite{OL_Cloak}. However, a single success \cite{OL_Cloak} is not sufficient to unambiguously prove that 3D STED-DLW is superior to the best regular 3D DLW. Thus, we now discuss a systematic comparison (regular DLW \textit{vs.} STED-DLW) on the ``drosophila'' of 3D direct laser writing, namely the woodpile photonic crystal \cite{3DPBG}. At this point, we also make the transition from summarizing previously published research to presenting original work. 

\section{Three-Dimensional Woodpile Photonic Crystals \textit{via} STED-DLW}
The 3D woodpile photonic crystal \cite{3DPBG} is composed of a first layer of rods evenly separated by the rod spacing $a$, a second similar but orthogonal layer on top, a third layer parallel to the first one but displaced laterally by half a rod spacing, and a fourth layer similarly displaced with respect to the second layer. These four layers result in a lattice constant $c$ in the axial direction. For $c/a=1$ one gets a body-centered-cubic (bcc) and for $c/a=\sqrt{2}$ a face-centered-cubic (fcc) 3D translation lattice. The woodpile is simple to write using 3D DLW and, at the same time, theoretically very well understood, especially regarding its optical spectra \cite{Deubel},\cite{WP_angle}. Furthermore, many groups have published corresponding results \cite{Misawa},\cite{Minggu},\cite{Deubel},\cite{WP_angle},\cite{PBG_telecom},\cite{Perry},\cite{cwDLW} allowing for a direct comparison. To the best of our knowledge, the smallest lateral rod spacings in fcc or bcc woodpiles published to date are $a=600\,\rm nm$ using near-infrared femtosecond pulses \cite{PBG_telecom}, $a=500\,\rm nm$ using 520-nm wavelength femtosecond pulses \cite{Perry}, and $a=450\,\rm nm$ using 532-nm continuous-wave exposure \cite{cwDLW}. 

\begin{figure}[htb]
\centering\includegraphics[width=13cm]{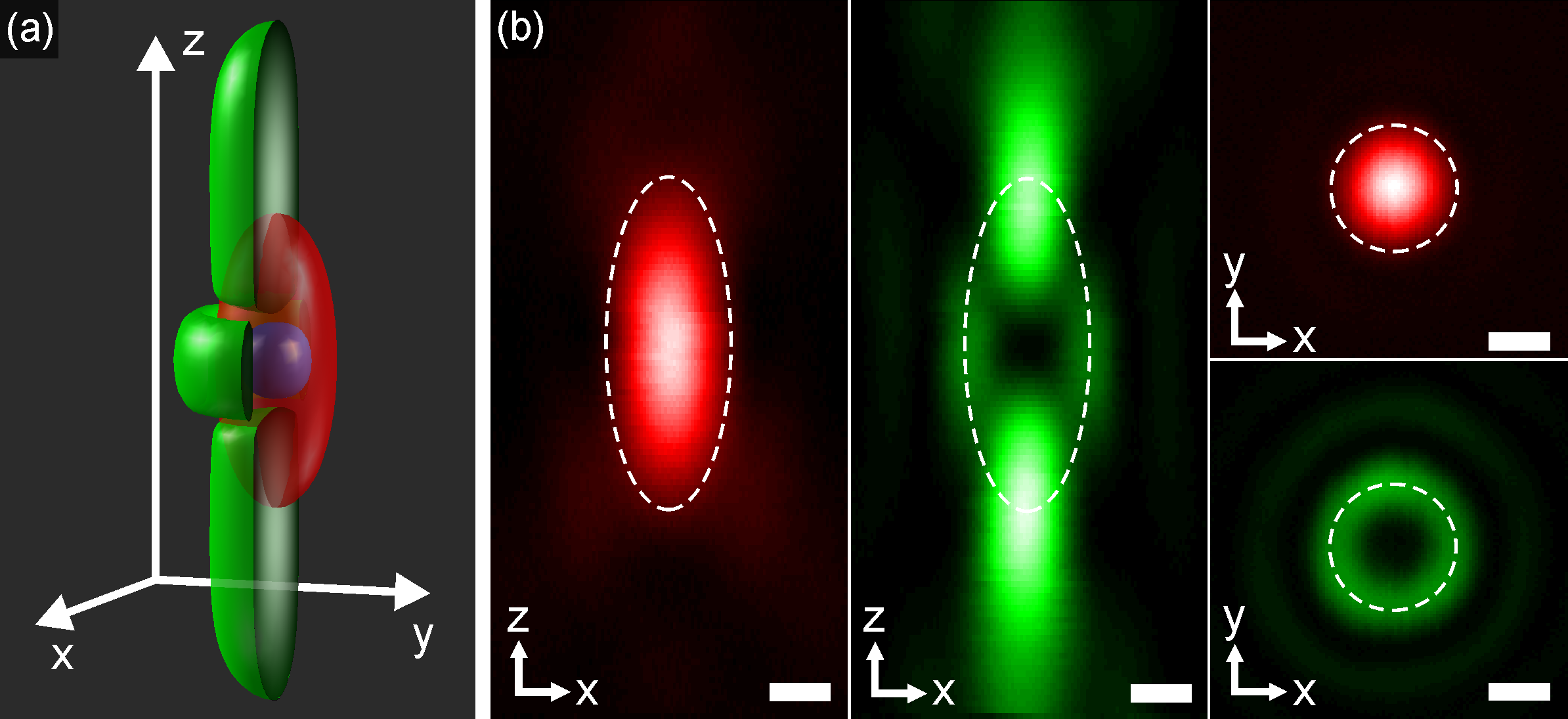}
\caption{(a) Calculated iso-intensity surfaces of the foci of the femtosecond excitation beam (red) and the continuous-wave depletion beam (green). This combination reduces the effective exposure volume (blue), both in axial (\textit{z}) and lateral (\textit{xy}) direction. (b) Measured intensity profiles of the two beams in the \textit{xz}-plane and in the focal plane (\textit{xy}). All scale bars correspond to 200\,nm.}
\end{figure}

In our present experiments, optical femtosecond pulses (Spectra Physics, Mai Tai HP) centered at around 810-nm wavelength and circularly polarized are tightly focused by a microscope objective (Leica HCX PL APO, ${\rm NA}=1.4)$ to polymerize a photoresist within the focal volume \textit{via} two-photon absorption. As shown in Fig. 2(a), a continuous-wave 532-nm wavelength (green) depletion beam (Spectra Physics, Millennia Xs) is shaped such that the elongated excitation focal volume (red) is effectively reduced to the desired more spherical exposure volume (blue). This shaping is accomplished by a home-made phase mask. This phasemask consists of a 430-nm high cylinder (photoresist SU-8, MicroChem) with refractive index $n=1.62$ at 532-nm wavelength on a glass substrate. This arrangement introduces a 180$^{\circ}$ phase shift in the center of the continuous-wave circularly polarized collimated beam. The phase-mask plane is imaged onto the entrance pupil of the microscope objective of the DLW system such that the central area occupies about 50\% of the entrance pupil area. The measured focal intensity distributions of the red and the green beam shown in Fig. 2(b) are obtained by scanning a single 100-nm diameter gold bead embedded in a monomer through the focus and recording the back-scattered light intensity. These intensity distributions show that the green depletion profile not only reduces the red profile along the axial \textit{z}-direction, but also comprises a ring in the focal \textit{xy}-plane, improving the lateral resolution as well. Corresponding foci are known from STED microscopy \cite{Hell2}, although less commonly used than donut depletion foci. 

\begin{figure}[htb]
\centering\includegraphics[width=13cm]{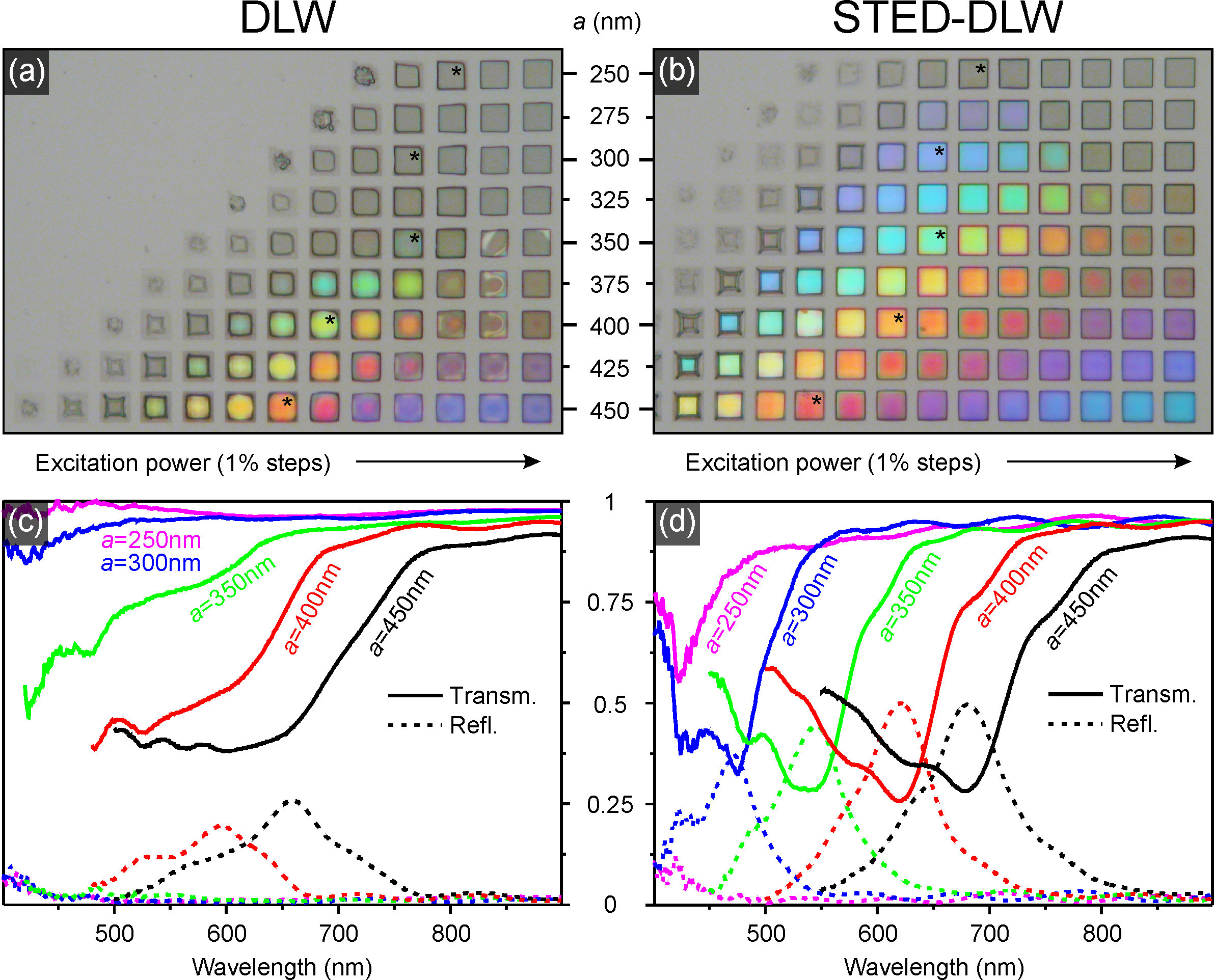}
\caption{(a) True-color reflection-mode optical micrographs of woodpile photonic crystals fabricated \textit{via} regular DLW. (b) Same, but using STED-DLW. All woodpiles have 24 layers and a footprint of $20\times20\,{\rm \upmu m}^2$. The rod spacing is decreased from $a=450\,\rm nm$ to $a=250\,\rm nm$ along the vertical, the exposure power is increased in steps of 1\% from left to right. (c) and (d) Selected (see asterisks in (a) and (b)) transmittance (solid) and reflectance (dashed) spectra for DLW and STED-DLW, respectively.}
\end{figure}

The photoresist used in our present work consists of 0.25\%\,wt DETC dissolved in the monomer pentaerythritol tetraacrylate \cite{OL_Cloak} (see Fig. 1(b)). The latter contains 300-400 ppm monomethyl ether hydroquinone as inhibitor. For all structures a constant scan velocity of $100\,\upmu \rm m/s$ is used. In order to reduce the unwanted resist exposure by the depletion beam, both beams are chopped with 3\% duty cycle at 4\,kHz frequency using acousto-optic modulators (also for the experiments without depletion beam). While maintaining a high depletion power during the modulators' on-state (corresponding to a high depletion efficiency), the overall absorption of the depletion beam by photoiniatior molecules in their ground state is greatly reduced due to the small duty cycle. Hence, the polymerization threshold for the green beam shifts towards larger powers. After DLW or STED-DLW, all photoresist structures are developed in 2-propanol and rinsed in acetone and water.

Fig. 3(a) and (b) show true-color optical micrographs (reflection mode) of woodpiles with lateral rod spacings ranging from $a=450\,\rm nm$ down to $a=250\,\rm nm$, each with $20\,{\rm \upmu m} \times 20{\rm \upmu m}$ footprint and 24 layers in the axial direction. All woodpile samples shown in Fig. 3 have been written with an axial lattice constant \textit{c} corresponding to an fcc lattice, {\it i.e.}, $c/a=\sqrt{2}$. To compensate for the anticipated shrinkage, the \textit{c}/\textit{a} ratio is further increased by 28\% in the writing process. The structures in Fig. 3(a) have been fabricated \textit{via} regular DLW (\textit{i.e.}, depletion laser switched off), those in Fig. 3(b) \textit{via} STED-DLW with 50-mW continuous-wave depletion power. In each row, the excitation power has been varied in steps of 1\% of the optimal power. The optimum excitation powers for regular DLW range from 7.4\,mW to 8.3\,mW for rod spacings from $a=250\,\rm nm$ to $a=450\,\rm nm$, respectively. The optimum excitation powers for STED-DLW are 31\% higher. All powers are quoted in front of the microscope-objective-lens' entrance pupil. The characterization measurements shown in Fig. 3 have been performed a few days after fabrication. At this point, the samples made by regular DLW have degraded to some extent, possibly due to insufficient cross-linking, whereas the samples made by STED-DLW have not shown significant degradation. 

The structures fabricated \textit{via} STED-DLW (Fig. 3(b)) clearly show brighter colors and are more homogeneous  compared to those made by regular DLW (Fig. 3(a), see woodpile edges). These colors result from Bragg-reflection off of the periodic structure. For a given rod spacing \textit{a}, the excitation-power window that leads to colorful, hence open, structures is much larger using STED-DLW compared to regular DLW. Obviously, STED-DLW also allows for significantly smaller rod distances while maintaining functionality.

Fig. 3(c) and (d) show transmittance and reflectance spectra recorded using a Fourier-transform microscope-spectrometer for unpolarized light and normal incidence (for details see \cite{Deubel}). Fig. 3(d) shows pronounced stop bands evidencing excellent sample quality and confirms the above qualitative findings. In particular, using STED-DLW, even the woodpiles with $a=250\,\rm nm$ show indications for a stop band at around 400-nm wavelength, approaching the ultraviolet. Notably, woodpiles with $a=350\,\rm nm$ and $a=250\,\rm nm$ were only recently achieved by a group from Sandia National Laboratory using state-of-the-art electron-beam lithography in a time-consuming layer-by-layer fashion \cite{ebeamWP}, leading to only 4 and 9 layers, respectively.

Obviously, we still have not broken the lateral Abbe criterion of 205\,nm in terms of lateral rod spacings here (see discussion in Section 1). However, we have significantly broken the axial criterion of 510\,nm. The closest axial center-to-center distance in a woodpile photonic crystal equals $\frac{3}{4}c$. This can be understood as follows. The second layer of rods overlaps with the first layer, the third layer is laterally shifted. The fourth layer contains voxels that are directly above voxels of the first layer. The distance between these two layers is $\frac{3}{4}c$. Clearly, the 28\% enlarged \textit{c} (used for shrinkage precompensation) has to be taken at this point, as shrinkage should not be misinterpreted as increased resolution. For regular DLW, the smallest woodpiles achievable in our experiments have $a=375\,\rm nm$, resulting in $d_{\rm axial}=\frac{3}{4}c=\frac{3}{4}\cdot \sqrt{2}a\cdot 1.28=509\,\rm nm$, which is pretty close to the axial Abbe criterion. In case of STED-DLW, woodpiles with $a=275\,\rm nm$ still show good quality, resulting in $d_{\rm axial}=\frac{3}{4}\cdot \sqrt{2}a\cdot 1.28=373\,\rm nm$ which is -- for the first time -- significantly below the Abbe limit.

\begin{figure}[htb]
\centering\includegraphics[width=13cm]{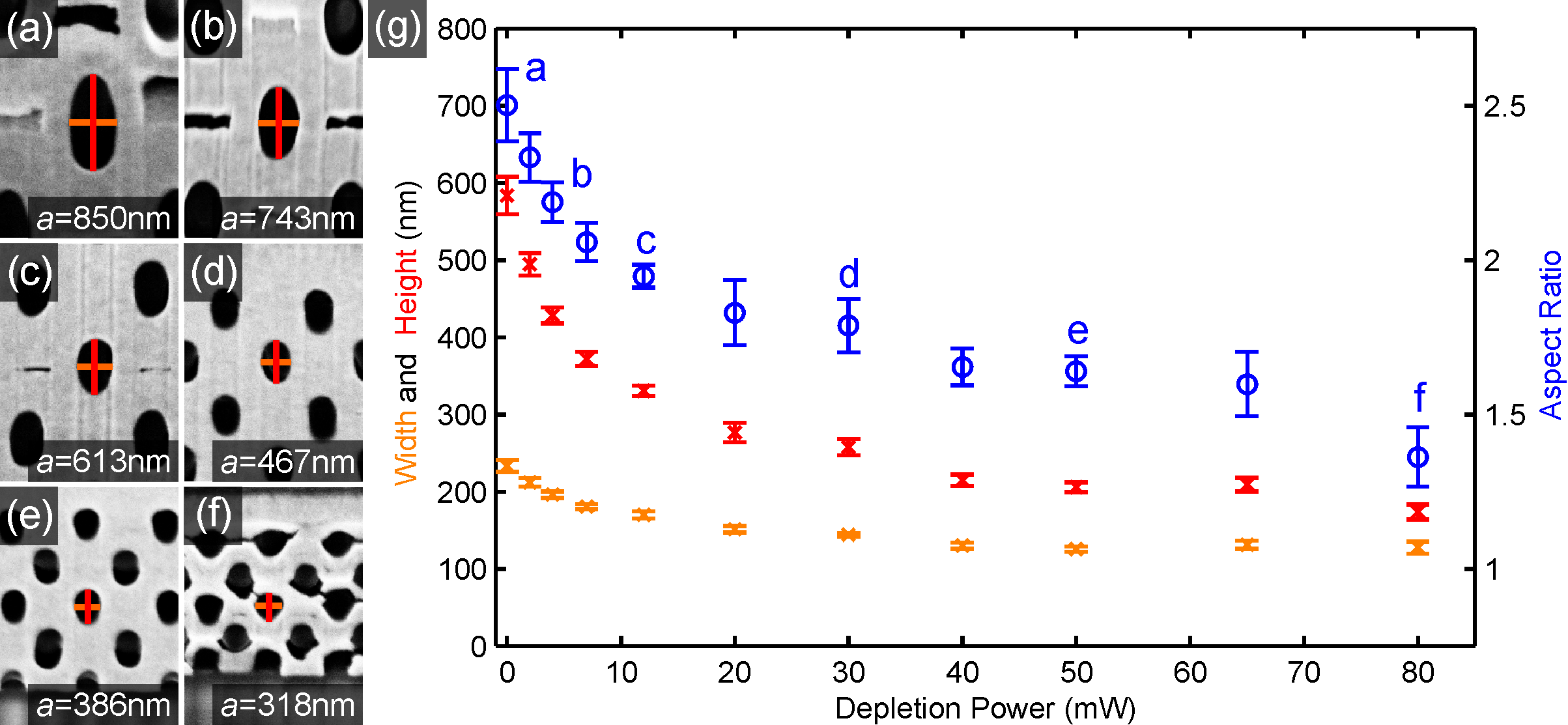}
\caption{(a)-(f) Oblique-view electron micrographs of ZnO-filled woodpile photonic crystals after focused-ion-beam milling. The viewing angle with respect to the surface normal is 54$^{\circ}$. (g) Width, height and calculated aspect ratio of polymer rods inside the three-dimensional woodpiles ((a)-(f)). Height measurements have been corrected for the viewing angle. The measurements are averaged over 10 rods. The error bars indicate $\pm$ one standard deviation of the corresponding ensembles. The bars for height and width in (a)-(f) correspond to the averaged values shown in (g).}
\end{figure}

In order to study this improvement in axial resolution and the anticipated reduction of the aspect ratio in more detail, we have fabricated another series of woodpile photonic crystals with a fixed excitation power and incrementally increasing depletion power. As the size of the rod cross-sections is supposed to decrease in all directions, we have adjusted the rod spacing $a$ accordingly to end up with connected structures with comparable filling fractions for all depletion powers. Characterizing the rods of photonic crystals instead of characterizing single voxels or lines close to the substrate surface has several advantages. First, potential (optical or diffusion kinetic) influences of the interface are eliminated. Second, shrinkage-dominated effects can be excluded, as strongly shrinking lines would not be able to form a complex three-dimensional structure, but would simply collapse. In order to assure that focused-ion-beam milling and electron-beam exposure do not alter the shape of the rods, we have filled the air gaps inside of the polymer structures with ZnO \textit{via} standard atomic-layer deposition (Cambridge NanoTech Inc.). Furhermore, ZnO is sufficiently conducting to eliminate the need for gold coating before scanning electron microscopy.

Fig. 4(a)-(f) shows oblique-view electron micrographs of these composite structures after focused-ion-beam milling. The bright material is ZnO, the dark areas correspond to the initial shape of the polymer rods. The widths and heights of the cross-sections have been measured for 10 rods per structure. The height measurements have, of course, been corrected for the viewing angle (54$^{\circ}$ with respect to the surface normal) and corresponding values for the aspect ratio (height/width) have been calculated. Averages of the resulting values are plotted versus depletion power in Fig. 4(g). Clearly, both the width and heigth of the features decreases with increasing depletion power. As expected from the depletion mode used, the decrease in height is more pronounced than that in width. Thus, the aspect ratio decreases from 2.5 at zero depletion power to a level of about 1.6 at larger depletion powers. The further decrease for 80\,mW depletion power is probably influenced by increased shrinkage, as the decreasing overall exposure dose approaches the polymerization threshold at this point.

\begin{figure}[htb]
\centering\includegraphics[width=13cm]{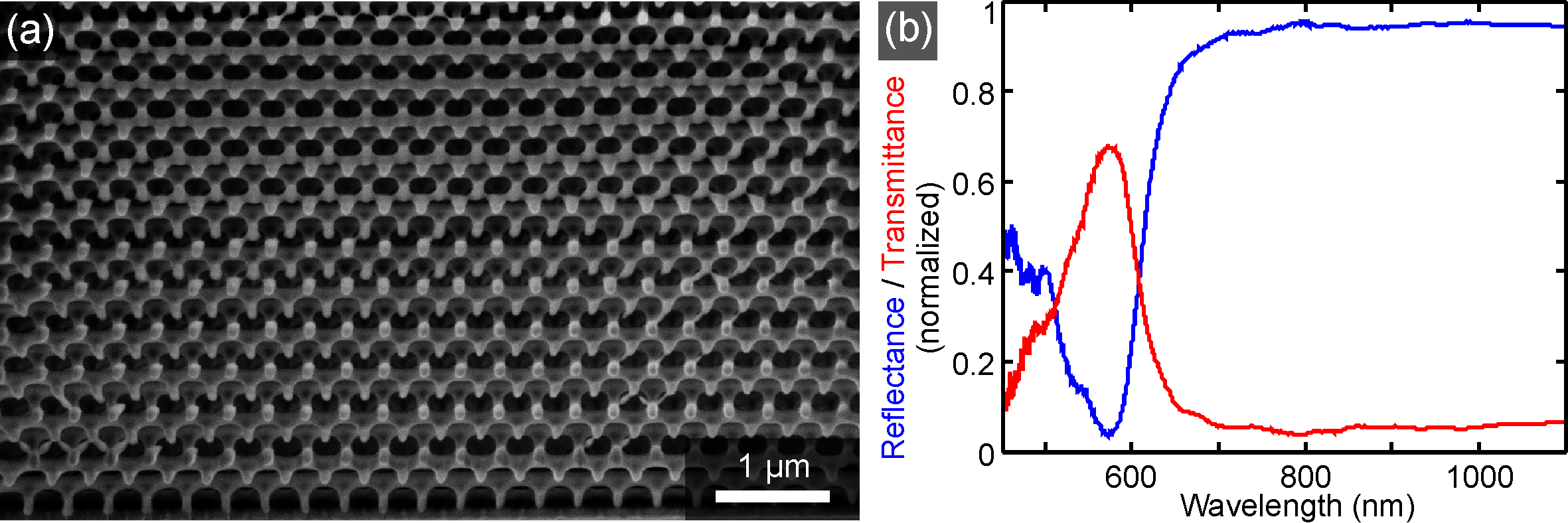}
\caption{(a) Oblique-view electron micrograph of a woodpile photonic crystal with 52 layers and a rod spacing of $a=350\,\rm nm$ made by STED-DLW. The sample has been milled with a focused-ion-beam to reveal its interior. (b) Corresponding reflectance and transmitted spectra (normalized to substrate transmittance and the reflectance of an 80-nm silver film, respectively).}
\end{figure}

One might fear that optical aberrations (due to the small but finite refractive-index contrast between photoresist and glass substrate as well as between polymerized and unpolymerized photoresist regions) could forbid the fabrication of very many layers in the axial direction in the case of STED-DLW, where both the excitation and the depletion focus are subject to aberrations. Fig. 5(a) shows a side-view electron micrograph of a woodpile with 52 layers (corresponding to a total height of $6.4\, \upmu\rm m$) and $a=350\,\rm nm$ made by STED-DLW after subsequent focused-ion-beam milling. Again, the depletion power has been 50\,mW. Obviously, the quality of the sample interior is excellent over its entire thickness. Fig. 5(b) shows the corresponding optical spectra with a pronounced stop band and a minimum transmittance of 3.5\%.

\section{Line Gratings}
Finally, we briefly address the current lateral resolution of STED-DLW and our novel photoresist in a separate set of experiments. Here we use a donut depletion mode like in our previous work \cite{STEDDLW} and write simple line gratings with variable period $a$. Corresponding examples from the literature demonstrate center-to-center distances $a=300\,\rm nm$ using 780-nm femtosecond pulses \cite{quencherpaper} or 532-nm continuous-wave exposure \cite{cwDLW}. As discussed in Section 1, this value is significantly above the respective diffraction-limited distances of $d_{\rm lateral}=197\,\rm nm$ (for $\lambda=780\,\rm nm$ \cite{quencherpaper}) and $d_{\rm lateral}=135\,\rm nm$ (for $\lambda=532\,\rm nm$ \cite{cwDLW}, assuming two-photon-absorption of the continuous-wave excitation). 

\begin{figure}[htb]
\centering\includegraphics[width=13cm]{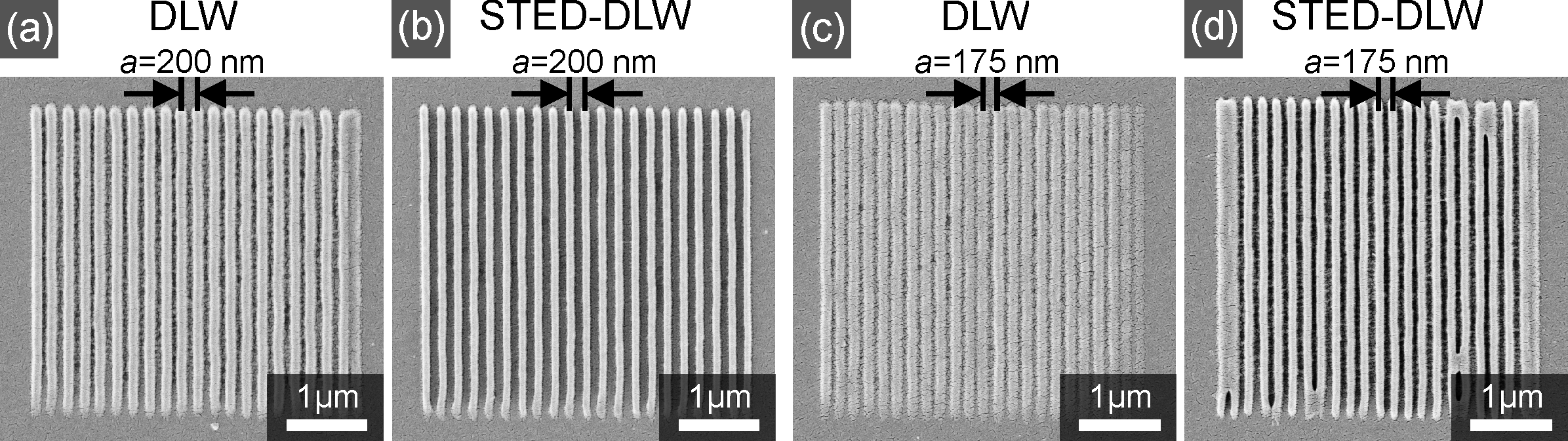}
\caption{Electron micrographs of simple line gratings fabricated \textit{via} regular DLW ((a) and (c)) and STED-DLW ((b) and (d)). The center-to-center distance of the lines is $a=200\,\rm nm$ and $a=175\,\rm nm$ as indicated within the panels. The depletion power of the donut mode used is 50\,mW in front of the microscope-objective-lens entrance pupil.}
\end{figure}

In our experiments we vary the excitation power, the depletion power, and the \textit{z}-position of the focus with respect to the substrate-resist interface. Fig.\,6 shows corresponding best results for regular DLW and for STED-DLW. For $a=200\,\rm nm$ (right at the diffraction limit), the grating fabricated by regular DLW (Fig. 6(a)) is hardly separated and polymer clusters bridge the gaps between the lines. These clusters result from dose accumulation due to the adjacent line exposures. Using STED-DLW, we get a grating of excellent quality with clean spacings (Fig. 6(b)). For $a=175\,\rm nm$, regular DLW does not allow for the fabrication of gratings with lines that are clearly separated from the substrate. Chosing the \textit{z}-position of the focus such that the major fraction of its axial profile is located within the substrate, we only obtain a periodic (yet very flat) height variation (Fig. 6(c)). In contrast, using STED-DLW (Fig. 6(d)), elevated and simultaneously separated features are still possible with reasonable quality. Obviously, this structure is slightly below the two-photon Abbe criterion of $d_{\rm lateral}\approx 205\,\rm nm$.

\section{Conclusion}
We have presented our early results on improving the resolution of three-dimensional direct laser writing by combining it with the concept of stimulated emission depletion known from fluorescence microscopy. The axial resolution exceeds the Abbe diffraction-barrier for the first time. Lateral resolution improvement is obtained as well, for the first time slightly exceeding the lateral Abbe criterion. These results raise hopes that diffraction-unlimited optical lithography may truly become the 3D counterpart of 2D electron-beam lithography, which has served as a workhorse for the entire field of (planar) nanotechnology for many years already. However, substantial further photoresist research is likely necessary to actually realize minimum feature sizes in the range of 10-30\,nm.

\paragraph*{\bf Acknowledgements}
We thank Andreas Fr\"olich for the ZnO ALD, Georg von Freymann and Tolga Ergin for helpful discussions. We acknowledge support through the DFG Center for Functional Nanostructures (CFN) within subproject A1.4. The project PHOME acknowledges the financial support of the Future and Emerging Technologies (FET) programme within the Seventh Framework Programme for Research of the European Commission, under FET-Open grant number 213390. The project METAMAT is supported by the Bundesministerium f\"ur Bildung und Forschung (BMBF). The PhD education of J.F. is embedded in the Karlsruhe School of Optics \& Photonics (KSOP).

\end{document}